\documentclass[12pt]{article}
\usepackage[super]{cite}
\usepackage{graphicx}
\usepackage{subfigure}
\usepackage{epsfig}
\usepackage{amsmath}
\usepackage{amssymb}
\usepackage{latexsym}
\usepackage{float}
\def\be{\begin{equation}}
\def\ee{\end{equation}}
\def\beq{\begin{eqnarray}}
\def\eeq{\end{eqnarray}}
\def\bn{\begin{eqnarray*}}
\def\en{\end{eqnarray*}}

\def\w{\omega}

\def\a{\alpha}
\def\b{\beta}

\def\d{\delta}
\def\D{\Delta}

\def\t{\theta}

\def\pd{\partial}

\def\k{\kappa}

\def\t{\theta}

\def\l{\lambda}

\def\cM{{\cal{M}}}

\def\cR{{\cal{R}}}

\def\cF{{\cal{F}}}

\def\bsp{\beq\begin{split}}
\allowdisplaybreaks
\mathchardef\mhyphen="2D
\def\bw {\bar{w}}

\begin{document}
\centerline{\bf Modelling quantum black hole}
\begin{center}
T. R. Govindarajan\footnote{trg@cmi.ac.in}\\
Chennai Mathematical Institute, H1, SIPCOT IT Park,
Kelambakkam, Siruseri 603103, India\\
\vskip3mm
J.M. Mun\~oz-Casta\~neda\footnote
{jose.munoz.castaneda@upm.es} \\
Physics Department, ETSIAE
Universidad Polit\'ecnica de Madrid,Spain
\end{center}
%%%%%%%%%%%%%%%%%%%%%%%%%%%%%%%%%%%%%%%%%%%

\date{\today}

\begin{abstract}
Novel bound states are obtained for manifolds with singular 
potentials. These singular potentials require proper boundary 
conditions across boundaries. The number of bound states match nicely
with what we would expect for blackholes. Also they serve to model 
membrane mechanism for the blackhole horizons in simpler contexts. 
The singular potentials
can also mimic expanding boundaries elegantly, there by obtaining 
appropriately tuned radiation rates. 
%\keywords{Theory of Quantized Fields; Field Theory; Edge and boundary effects.}

\end{abstract}
%\maketitle
\section{Introduction} W. Pauli {\cite{pauli}}remarked the 
boundaries were 
the creation of the devil. 
Bekenstein's area law {\cite{bekenstein,hawking}} for the entropy 
of blackhole prescribes that the microscopic states live close to the
horizon and the number of such states grow rapidly with area. 
One can proceed at least in the case 
of large blackholes without the detailed requirements of quantum 
geometry to study quantum blackholes. Such attempts 
have been made earlier by 't Hooft through 
a brick wall {\cite{thooft}}, and Beckenstein and 
Mukhanov{\cite{mukhanov}}. The entropy
is understood in some context by entanglement of those inside the 
horizon which are inaccessible to asymptotic observer with the 
outside{\cite{sorkin,srednicki}}. 

In this communication we elaborate an earlier proposal{\cite{trg1}} 
by one of us that the 
existence of bound states in the blackhole geometries resulting from
the study of self adjoint extensions of the Laplacian near the 
horizon. Near horizon geometry of blackholes 
present a singular potential to
the particles and can be studied through quantum mechanics 
with special boundary conditions. Not only they lead to localised 
states  on the boundary their number also scales like area. This in
statistical mechanics or in quantum field theory (QFT) 
context translates as entropy 
\cite{wheeler}. Similar is the case of von Neumann entropy 
when we trace over unobserved bound states for a distant observer.    
%%%%%%%%%%%%%%%%%%%%%%%%%%%%%%%%%%%%%%%%%%%%%%%
The quantum physics on manifolds 
with boundaries introduces novel features. They appear in varied 
situations  like Casimir effect {\cite{casimir,casimir1,casimir2,
karabali}}, quantum hall, topological
insulators{\cite{ti,asorey-ti} or quantum gravity contexts 
like blackhole, 
or de Sitter spacetime with cosmological horizon {\cite{membrane,
damour}}. Many of the novel 
features stem from studying correct boundary conditions which 
are physically relevant as well mathematically correct to make 
the Hamiltonian a self adjoint operator in properly extended 
domains in the Hilbert space of $L^2$ functions. 

The garden variety boundary conditions are Dirichlet and Neumann
for which either the function or the normal derivative vanishes. 
However as shown for the Laplacian a more general class of 
boundary condition is possible, a particular example being 
Robin boundary condition. Here 
the function and the normal are related on the boundary $(\psi
~+~\k\pd \psi)|_{\pd \cM}~=~0$. Dirichlet and Neumann are extreme 
limits of the Robin boundary condition. But more importantly 
it introduces a fundamental length $\k$ 
into the theory{\cite{trg2,trg3}}. 
Such a parameter will emerge from the coarse grained structure 
of the underlying space-time or crystalline lattice. 
Typically it can be related to Planck 
length in a semiclassical gravity context. 
    
But the boundaries are obtained in reality 
through singular potentials or point interactions 
in space-time and such potentials 
are also subjected to self adjointness conditions  
on unbounded operators{\cite{reed, asorey,asoreyjose}. 
The well known potential of this kind 
in one dimension is the $\d(x)$ which introduces discontinuity 
in the derivative of the wavefunctions. More general 
potentials of the same type  is the $\d'(x)$ potential which has 
the new feature of introducing discontinuities in the 
wavefunction itself. Inspite of such discontinuities the Hamiltonian
remains  self adjoint and the quantum theory describes well defined
unitary evolution. This approach allows study of quntum fields over 
bounded regions in terms of interesting and meaningful questions that
can be answered. One can sacrifice the self adjointness with special 
boundary conditions like purely ingoing waves leading to quasi normal
modes (QNM) which are also linked to ringing modes of stellar objects 
including blackholes. \cite{kokkotas}.
          
We will consider quantum theory with point interactions of the type 
which is a combination of $\d$ and $\d'$ potentials. 
Such a combination
in addition to being more general, is also necessitated for several
reasons. They arise naturally when we consider self adjoint 
extensions of Dirac operator with singular $\d$ potential
{\cite{trg4}}. But for us 
it brings new features like what we anticipate from membrane 
paradigm{\cite{membrane,damour}}for black holes through a new parameter. 
Our constructions can easily be extended to curved backgrounds too.

Introduction of singular distributions as potentials also help 
in introducing time dependent boundaries and associated radiation.
We will study in this communication in Sec 2 
a model Schrodinger equation 
in $R^2$ wih singular point interaction potentials $a\d(r-R)~+~2b
\d'(r-R)$. We consider in Sec 2a the scattering states and in Sec 2b
the bound states. In Sec 2c we explain the extension to 3 dimensions.
We also remark about BTZ blackhole in this context. In Sec 3 we 
consider moving boundaries through singular potentials 
and explain how this 
program can be carried out{\cite{trgjose}}. 
\section{The model}
The general study of point interactions of the free 
Hamiltonian in one dimension $H~=~\frac{\hbar^2}{2m}\frac{d^2}{dx^2}$
is due to Kurasov {\cite{kurasov,kurasov1}} 
and uses von Neumann's theory of symmetric 
unbounded operators with identical deficiency indices{\cite{reed}. 
The general
analysis of self adjointness of Laplacian in higher dimensional
manifolds with boundaries is more complex due to infinite 
deficiency indices. But it is possible to relate them directly to 
boundary conditions of functions and normal derivatives on the 
boundaries{\cite{asorey,asoreyjose}}. 
In cases like ours, the presence of isometries  
simplifies the problem considerably and provide exact solutions 
{\cite{trg2}}. 

Consider a Schrodinger Hamiltonian equation in $R^2$ for stationary 
states with a singular potential along a circle of radius $R$:
\be
\left[-\frac{\hbar^2}{2m}\D+a\d(r-R)+b\d'(r-R)\right]
\Psi(r,\t)=E\Psi(r,\t)
\label{1_1}
\ee
In order to work with dimensionless quantities, let us introduce
new variables and parameters.
\begin{equation}
\label{1_2}
{\bf r}=\frac{\hbar}{mc}\, {\bf x},\
R~=~\frac{\hbar}{mc}\, X,\
w_0= \frac{2a}{\hbar c},\
w_1= \frac{mb}{\hbar^2},\
\l= \frac{2E}{m c^2}, \
\end{equation}
such that \eqref{1_1} becomes with $\varphi({\bf x})=\Psi(r,\t)$,
\begin{equation}
\label{1_3}
-\Delta_{\bf x}\,\varphi({\bf x}) +
w_0\delta(x- X) \varphi({\bf x}) +
2 w_1 \delta'(x- X) \varphi({\bf x})
=\l\, \varphi({\bf x}).
\end{equation}
This new parametrization corresponds to lengths being measured 
in the units of Compton wavelength of the particle.  
The origin of $a$ is related to the underlying background
geometry and is independent of `m'. On the other hand `b' is 
related to the mass `m'.  
 
The crucial question is to find the domain of
wave functions $\varphi({\bf x})$ that makes $H_0$ self adjoint.
As these
functions and their derivatives have a discontinuity at
$x=X$, we have to define the products of the form
$\delta(x-X) \varphi({\bf x})$ and
$\delta'(x-X) \varphi({\bf x})$ in \eqref{1_1}. The 
form for these products are given as:
\begin{eqnarray}
\label{1_222}
\delta(x-X) \varphi({\bf x})&=&\frac{\varphi(X^+,\t)+
\varphi({X}^-,\theta)}{2}\,\delta({x-X})\,,\nonumber \\  
\delta'({x-X}) \varphi({\bf x})&=&\frac{\varphi({X}^+,\t)+
\varphi({X}^-,\t)}{2}\,\delta'({x-X})\nonumber \\
&-&\frac{\varphi'({X}^+,\theta)+\varphi'({X}^-,\theta)}{2}\,
\delta({x-X})
\end{eqnarray}
where $f({X}^+,\theta)$ and $f({X}^-,\theta)$ are the 
right and left limits of the function $f({\bf x})$ as 
${x}\to {X}$, respectively. The problem is 
separable and can be reduced to a 1D radial problem with a 
central potential.
In order to obtain a self-adjoint extension of the Hamiltonian 
we have to find a domain on which this extension acts, namely given 
by a space of square integrable functions satisfying matching 
conditions at the point $X$. The radial functions in the 
domain of the Hamiltonian $H$ are functions in the Sobolev space
$W^2_2({\mathbb R}/\{S^2(X)\})$ such that at $x=X$ 
satisfy the following matching conditions given by an 
$SL(2,R)$ matrix{\cite{gadella}}:
\begin{equation}\label{1_4}
\left( \begin{array}{c}
\varphi(X^+) \\[1ex]
\varphi'(X^+)  \end{array}  \right) =
\left( \begin{array}{cc}
\displaystyle\frac{1+w_1}{1-w_1}  &  0
\\[1ex]
\displaystyle
\frac{  w_0}{1-w_1^2}  &  \displaystyle \frac{1-w_1}{1+w_1} \end{array}  \right) \left( \begin{array}{c}
\varphi(X^-)  \\[1ex] \varphi'(X^-)  \end{array}  \right)\,.
\end{equation}     
Note that in the case of $w_1$ being  zero it goes to 
known discontinuities
in normal derivatives {\cite{trg4}}.
%%%%%%%%%%%%%%%%%%%%%%%%%%%%%%%%%%%%%%%%%%%%%%%%%%%%%
\subsection{Scattering states}
%%%%%%%%%%%%%%%%%%%%%%%%%%%%%%%%%%%%%%%%%%%%%%%%%%%%%
For scattering theory we solve Schrodinger equation with plane waves
and positive energy $k^2$. For each angular momentum $n$
we obtain the following Schrödinger 1D problem for $\varphi(r,\t)
~=~\cR~e^{in\t}$ :
\beq
\frac{d^2 \cR}{dr^2}+\frac{1}{r}\frac{d\cR}{dr}&+&\left(w_0\delta(r-R)
+2w_1\d'(r-R)\right)\cR \nonumber \\
&-&\left(\l+\frac{n^2}{r^2}\right)\cR=0
\label{schrodinger}
\eeq
with $\l~=-~k^2$ and suitable boundary conditions. 
The general scattering solution is given by
\begin{equation}
\cR(r)=\begin{cases}
 J_n(k r) & r < R\\
 A(k,n)J_n(kr)+B(k,n) Y_n(k r) & r > R
\end{cases},
\end{equation}
where $J_n$ and $Y_n$ are the Bessel functions, 
$A(k,n)$ and $B(k,n)$ constants to be determined through matching 
boundary conditions.
\bn
B(k,n)&=&J_n(x)\left(4 k w_1~R~J_{n-1}
(x)-J_n(x) \left(4w_1n~+~
w_0R\right)\right)\nonumber \\
A(k,n)&=&J_n(x) \left(k (w_1 +1)^2 R
   Y_{n-1}(x)-Y_n(x) \left(4w_1L+
w_0R\right)\right)\nonumber\\
&-&k(w_1 -1)^2R~J_{n-1}(x)Y_n(x)
\en
where we have defined $x~=~kR$.
%%%%%%%%%%%%%%%%%%
This complicated looking expression can be checked to 
coincide with expected results for hard 
sphere ($w_1~=~-1$). Note that when $w_1=1$ the exterior side 
of the Disc $D_2$ is seen by the quantum particle as 
Robin boundary condition while the inside face is Dirichlet. 
On the other hand for $w_1=-1$ this is the other way round. 
The phase shifts are given by
$\tan(\delta_n)=-B(k,n)/A(k,n)${\cite{phaseshift} where $A,B$ 
are given above.
%%%%%%%%%%%%%%%%%%%%%%%%%%%%%%%%%%%%%%%%%%%%%%%%%%%%%%%%%%%%%%%%%%
\subsection{Bound states}
%%%%%%%%%%%%%%%%%%%%%%%%%%%%%%%%%%%%%%%%%%%%%%%%%%%%%%%%%%%%%%%%%
The Schrodinger equation is Eq. ({\ref{schrodinger}}) with
$\l~\geq~0$. 
The solutions in the regions $r~<~R ~and~ r~>~R$ are the modified 
Bessel functions of the first and second kind: $\cR(r)~=~
c~I_n(\sqrt{\l}r)$~and~$d~K_n(\sqrt{\l}r)$. We match the boundary 
conditions at $r~=~R$ using Eq.{\ref{1_4}}.
%%%%%%%%%%%%%%%%%%%%%%%%%%%%%%%%%%%%%%%%%%%%%%%%%%%%%%%%%%%%%%%
We rewrite: $\a~=\frac{1+\w_1}{1-\w_1},~~\b~=~\frac{2\w_0}{1-w_1^2}$.
We get (with $x~=~\sqrt{\l}R$),
\be
x\left(\a\frac{K_n'(x)}{K_n(x)}~-~\a^{-1}\frac{I_n'(x)}{I_n(x)}
\right)~=~\bar{\b}
\ee
where $\bar{\b}~=~\b~R$.
We can simplify the above equation using Bessel function identities
to get:
\be
-x\left(\frac{\a~K_{n-1}}{K_n}~+~\frac{\a^{-1}I_{n-1}}{I_n}\right)
~-~n~(\a~-~\a^{-1})=~\bar{\b}
\label{boundstates}
\ee
Now we can look for a maximum value of 
$n~=~n_m$. It is easy to work out:
This gives:
\be
-~n_m~(\a~+~\a^{-1})~=~\bar{\b}
\ee
Hence the maximum number of bound states are still proportional 
to the radius of the circle, but with a renormalised 
constant $\frac{w_0}{2(1~+~w_1^2)}$. 
For the special case of $w_1~=~0$ we get maximum number $n_m$ of 
bound states is the nearest integer lower than $\bw_0$, which is same 
as our earlier result{\cite{trg1,trg2}}.
For the case when $w_1$ is small when we can drop
$w_1^2$ term we get
the number of bound states is unaltered. This is to be expected
since the singular potential can be written as shifted singular
potential: $V(r)~\approx~w_0~\d(r-R~+~\frac{2w_1}{w_0})$.

Energy eigenvalues are obtained numerically for different 
values $R, w_0$ and $w_1$ by solving Eq.(\ref{boundstates}). 
Similarly we obtain expectation values $<r_n>$ by using appropriate 
$\cR(r)$ in the two regions.   
The graphs demonstrate the number of bound states as well as energy 
eigenvalues (Fig.{\ref{fig1}}). They also explicitely show that 
they are localised close to the boundary and  deviate externally 
or internally when we increase the coefficent of $\d'$ potential
(Fig.{\ref{fig2},\ref{fig3}}). Interestingly the states 
are localised outside 
(inside) the boundary for positive (negative) $w_1$ respectively. 
$w_1$ can be tuned to reduce the probability of finding 
the particle inside model blackhole to be small. 
Also note that higher angular momenta states move closer 
to (away from ) the boundary for negative (positive) 
$w_1$. We will remark about this in the conclusions.  
%%%%%%%%%%%%%%%%%%%%%%%%%%%%%%%%%%%%%%%%%%%%%%%%%%%%%%%%%%%
\begin{figure}
\begin{minipage}[b]{.5\textwidth}
\includegraphics[width=\textwidth]{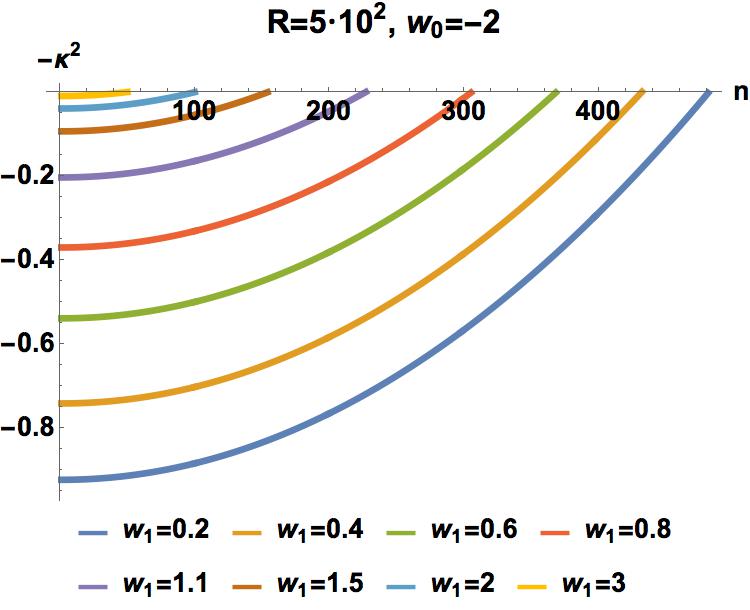}
\caption{\small Bound states $E(w_1)$
\label{fig1}}
\end{minipage}
%%%%%%%%%%%%%%%%%%%%%%%%%%%%%%%%%%%%%%%%%%%%%%%%%%%%%%%%%%%
\begin{minipage}[b]{.5\textwidth}
\includegraphics[width=\textwidth]{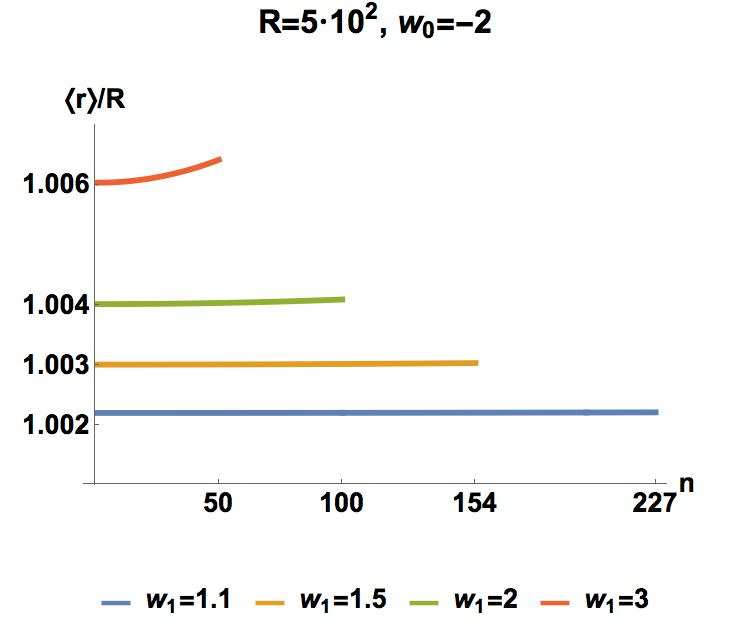}
\caption{\small $<r_n>/R$ for large $w_1$ 
\label{fig2}}
\end{minipage}
\end{figure}
%%%%%%%%%%%%%%%%%%%%%%%%%%%%%%%%%%%%%%%%%%%%%%%%%%%%%%%%%%%
\begin{figure}
\includegraphics[width=.7\textwidth]{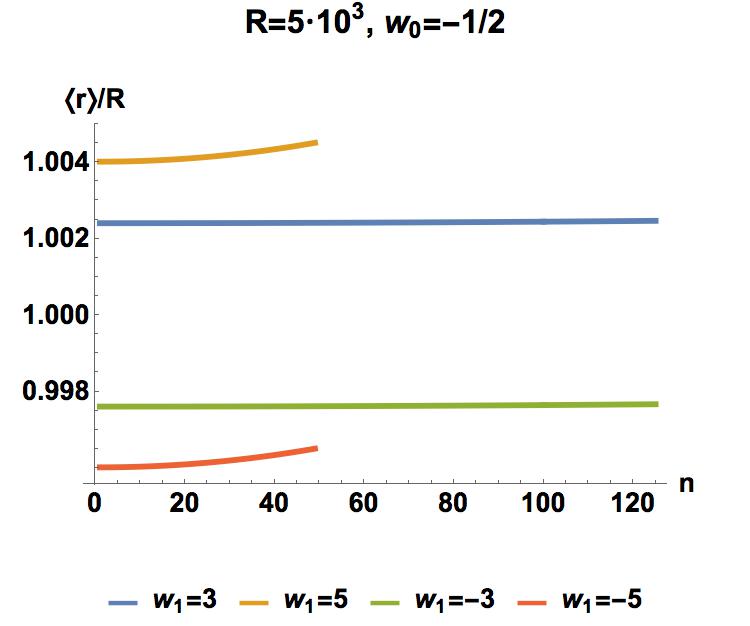}
\caption{\small Expectation $<r_n>/R$ for large R
\label{fig3}}
\end{figure}
%%%%%%%%%%%%%%%%%%%%%%%%%%%%%%%%%%%%%%%%%%%%%%%%%%%%%%%%%%
%%%%%%%%%%%%%%%%%%%%%%%%%%%%%%%%%%%%%%%%%%%%%%%%%%%%%%%%%%%
\subsection{Three dimensions and BTZ blackhole:}
We present in this section the results for bound states in $d~=~3$. 
For that we consider Schrodinger equation on $R^3$ with a singular 
potential along a sphere: $V(r)~=~a\d(r-R)~+~2b~\d'(r-R)$. 
The required Schrodinger equation in spherical polar coordinates
has solutions $\varphi(r,\t,\phi)~=~\cR(r)Y_{lm}(\t,\phi)$ where
$Y_{lm}$ are spherical harmonics solving angular 
part of the equations.
The radial part of the equation is:
\be 
\frac{d^2\cR}{dr^2}~+~\frac{2}{r}\frac{d\cR}{dr}~-~\frac{l(l+1)\cR}{r^2}
~=~\l \cR
\ee
The solutions in the regions $r~<~R ~and~ r~>~R$ are modified 
spherical Bessel functions: $\cR(r)~=~
c~\frac{I_{l+\frac{1}{2}}(\sqrt{\l}r)}{\sqrt{r}}$~and~ 
$d~\frac{K_{l+\frac{1}{2}}(\sqrt{\l}r)}{\sqrt{r}}$. 
Again matching  the boundary
conditions at $r~=~R$ and using Eq.{\ref{1_4}},
the Eq.({\ref{boundstates}) gets modified to:
\be
-x\left(\frac{\a K_{l-\frac{1}{2}}}{K_{l+\frac{1}{2}}}
+\frac{\a^{-1}I_{l-\frac{1}{2}}}
{I_{l+\frac{1}{2}}}\right)
-(l+1)(\a-\a^{-1})=\bar{\b}
\ee
The maximum angular momentum allowed $l_{max}~=~\frac{\bar{\b}+\a}
{\a~+~\a^{-1}}$.
Each angular momentum $l$ has degeneracy of $2l~+~1$ states.
Hence the number of states upto $l_{max}~\propto
2l_{max}^2$ we get the number of bound states $\propto R^2$.
%%%%%%%%%%%%%%%%%%%%%%%%%%%%%%%%%%%%%%%%%%%%%%%%%%%%%%%%%%%%%%%%

We will briefly mention the toy model of blackhole in 3d, BTZ 
blackhole {\cite{btz,btz1}}
For simplicity we consider BTZ blackhole metric with angular
momentum $J~=~0$. The metric is given by:
\be
ds^2=-\left(\frac{r^2}{l^2}-M\right)dt^2+
\left(\frac{r^2}{l^2}-M\right)^{-1}dr^2+r^2d\phi^2
\ee
Here cosmological constant $\frac{1}{l^2}$ and $M$ is the mass 
of the blackhole and horizon is at $r~=~r_+~=~\sqrt{M}l$. 
The solution of the scalar field in the presence of this metric 
along with singular potentials $\d(r-r_+)$ and $\d'(r-r_+)$
can be written as: $e^{-i\w t~+im\phi}\cR(r)$. Defining, 
$z~=~\frac{r^2~-~r_+^2}{r^2},~~~~\cF(z)~=~z^{i\a}(1-z)^{-\b}\cR(z)
$
one gets the hypergeometric differential equation for $\cF(z)$:
\be
z(1-z)\frac{d^2\cF}{dz^2}~+~(c-~(1+a+b))\frac{d\cF}{dz}~+~
ab\cF~=~0
\ee
where $a,b,c,\a,\b$ are constants defined terms of $r_+, l, m, \w$.
This hypergeometric equation has singularities at $z = 0,1$.
To obtain the bound states one should match the boundary 
conditions at $z=0$ for the hypergeometric functions
$\cF(a,b,c,z)$. 
%%%%%%%%%%%%%%%%%%%%%%%%%%%%%%%%%%%%%%%%%%%%%%%%%%%
\section{Expanding boundaries}  
%%%%%%%%%%%%%%%%%%%%%%%%%%%%%%%%%%%%%%%%%%%%%%%%%%%
In this brief section we explore the singular potentials 
for introduction of time dependence in boundaries.   
The simple case of in one dimension with $x~>~0$:
If the boundary is moving
with uniform velocity like $x~=~vt$ it can be studied as a quntum
mechanical problem with delta function potential $\delta(x-vt)$.
The solution is easy to get as $\psi(x,t)~\propto~e^{-|\kappa(x-vt)|}
e^{-i(\kappa^2-v^2/2)t-vx}$. 

We can easily extend this analysis to a boundary with an acceleration
`g' with a singular potential $\delta(x- \frac{gt^2}{2})$.
This is unitarily equivalent to the static singular potential
and an additional gravitional potential $mgx$.
This can be seen by using the unitary transformations:
$\phi(x,t)~=~U~V~\psi(x,t)$ where $V(x,t)~=~e^{-i\frac{g^2t^2p_x}{2}},
~~U(x,t)~=~e^{igxt~+~i\frac{g^2t^3}{6}}$
The solutions for linear gravitational potential
are given by Airy functions.

Similarly we can consider $R^2~-~D$. If the disc is expanding
it is better to convert the question to a delta function
potential which is expanding. 

Berry and Klein {\cite{berry}} showed the time dependent 
\begin{equation}
H(r,p,l(t))=\frac{p^2}{2m}+\frac{1}{l^2}V(r/l).
\end{equation}
can be simplified if the time dependence is of the form 
$l(t)=\sqrt{at^{2}+2bt+c}$. It becomes in a comoving frame:
\begin{equation}
H(\rho,\pi,k)=\frac{\pi^2}{2m}+V(\rho)+\frac{1}{2}k\rho^{2},
\end{equation}
where $\rho=r/l$ and $k=m(ac-b^{2})$ which is conserved
in $\rho,\tau\equiv\int^{t}\frac{dt}{l^{2}(t)}$.
The expanding disc in $\mathbb{R}^2$ and ball
in $\mathbb{R}^3$ will come under this class of Hamiltonians.
Consider the Hamiltonian {\cite{trg2}} in $R^2$
\be
H~=~ -\Delta ~+~ g~\delta(r~-~e^fR(0))~
\ee
By rescaling r we get the potential
as $e^f\delta(r~-~R(0))$. The
time dependence is shifted to the strength of potential. This is
analogous to changing the Hamiltonian to a time dependent
one by keeping the domain of the wavefunctions in 
the Hilbert space same for all times.

Applying Berry, Klein transformation {\cite{berry}}
we can convert the problem
in a comoving frame to a time independent potential with a delta
function along a ring. This will also correspond to generalised 
pantographic change of Fabio Anza etal {\cite{fabio}} 
This has important consequesnces for the rate of
emission or in expanding statistical ensembles with new boundary
conditions.
%%%%%%%%%%%%%%%%%%%%%%%%%%%%%%%%%%%%%%%%%%%%%%%%%%%%%%%%%%%%%%
\section{Conclusions}
%%%%%%%%%%%%%%%%%%%%%%%%%%%%%%%%%%%%%%%%%%%%%%%%%%%%%%%%%
In this report we have approached the question of 
quantum blackhole through
straightforward analysis of quantum theory on manifolds with 
boundaries or equivalently singular potentials. 
While our study is in Euclidean space it can be applied to curved 
background also since point interactions are local. This can
parallel the recent approach to understand blackholes through 
conventional notions of particles and 
forces treating blackholes just like atoms,
molecules (see 't Hooft{\cite{thooftnew}}). These require 
analysis through self adjoint extensions of operator domains. 
Our analysis surprisingly brings out the importance of both
$\delta$ and $\delta'$ potentials. There are  a number bound 
states localised  close to the boundary and is proportional to the 
area. As pointed out in the Introduction \cite{wheeler}
they relate to entropy in QFT.
Hence the existence of correct behaviour of localised bound states  
on the boundary is 
a strong requirement for correct entropy. We also point out the 
role of $\delta'$ potential in extending the support of the bound
states to enhanced length scales to allow for the possibility 
of quantum effects beyond Planck length{\cite{rovelli}}. 

Following tHooft \cite{thooft} one can consider scalar
fields to vanish at a small distance away from the horizon.
That is $\phi(R+h)~=~0$. This is for small $h$ equivalent to
Robin boundary condition since by expanding around R we get
$\phi(R)~+~h\phi'(R)~\approx 0$. This  boundary condition
can also be obtained from $\delta$ function potential. Our potential
is a generalisation of the potential which adds another parameter
which allows the quantum effects to persist beyond the length
parameter $h$.  In Kruskal coordinates one avoids
the singularity of the metric at the horizon,
but contain two copies of the space time. This
is mimicked in our case of singular potentials
connecting  the two regions with suitable boundary conditions
to maintain unitarity. Our generalised brickwall
mechanism can be studied  to obtain all the thermodynamic properties.
Detailed analysis using these
boundary conditions for the thermodynamic behaviour
will be presented elsewhere with Rindler, BTZ and Schwardschild
background(under preparation).

These states are interestingly connected through spectrum generating 
algebra which is a sub algebra of the Schrodinger group. 
By tuning the strength of the $\d'$ potential one can control 
the tunnelling through the boundary.  
Lastly if we scale the radius to $\infty$ keeping number of 
bound states fixed ($\frac{w_0}{w_1^2} \rightarrow 0$) the states
become zero energy bound states and localised at the boundary 
and play signifcant role for asymptotic symmetries.
The connections to QNM which arise from purely 
ingoing modes is also intriguing. In addition the 
singular potentials can be time dependent to enable the analysis 
of expanding boundaries and associated radiation output. This 
study leads us to new avenues of exploration to situations where 
boundaries and boundary conditions are involved{\cite{trgjose}}.    

Acknowledgements:
TRG and JMMC thank Manuel Asorey,University of
Zaragoza. 
TRG acknowledges discussions with Rakesh Tibrewala. JMMC acknowledge the funding 
by Spanish Government (project MTM2014-57129-C2-1-P) 
and discussions with L. M. Nieto, M. Gadella, 
and K.Kirsten.

%%%%%%%%%%%%%%%%%%%%%%%%%%%%%%%%%%%%%%%%%%%%%%%%%%%%%%%%%%%%

\end{document}